\documentclass[12pt,a4paper,final]{iopart}

\usepackage{iopams} 
\usepackage[breaklinks=true,colorlinks=true,linkcolor=blue,urlcolor=blue,citecolor=blue]{hyperref}
\usepackage{graphicx}
\usepackage{caption}
\usepackage{dcolumn}
\usepackage{bm}
\usepackage{subfig} 
\usepackage{epstopdf}
\usepackage{url}
\usepackage{subfig}
\usepackage{lipsum, appendix}

\newcommand{\upperRomannumeral}[1]{\uppercase\expandafter{\romannumeral#1}}

\begin{document}

\title{Exploring Critical Collapse in the Semilinear Wave Equation using Space-Time Finite Elements}

\author{Hyun Lim$^{1}$, Matthew Anderson$^{2}$, Jung-Han Kimn$^{3}$}
\address{$^1$ Department of Physics and Astronomy, Brigham Young University, Provo, UT 84602, USA}
\address{$^2$ Center for Research in Extreme Scale Technologies, Indiana University, Bloomington, IN 47405, USA}
\address{$^3$ Department of Mathematics and Statistics, South Dakota State University, Brookings, SD 57007, USA}


\begin{abstract}
A fully implicit numerical approach based on the space-time finite element method is implemented for the semilinear wave equation in 1(space) + 1(time) and 2 + 1 dimensions to explore critical collapse and search for self-similar solutions. Previous work studied this behavior by exploring the threshold of singularity formation using time marching finite difference techniques while this work introduces an adaptive time parallel numerical method to the problem. 
The semilinear wave equation with a $p = 7$ term is examined in spherical symmetry. The impact of mesh refinement and the time additive Schwarz preconditioner in conjunction with 
Krylov Subspace Methods are examined. 

\end{abstract}

\vspace{2pc}

\section{Introduction}
The semilinear wave equation is a physically interesting system to investigate because of the emergence of singularities from smooth initial data in finite time. Reminiscent of critical behavior in black hole formation discovered by Choptuik~\cite{ChoptuikPRL1993}, the semilinear wave equation provides a laboratory in which to investigate the transition regions between singularity formation and search for critical behavior in a much simpler system than the Einstein equations. A number of past studies have done precisely this~\cite{IsenbergJMP2002, BizonNL2001, LieblingJMP2002, BizonNL2000, LieblingPRAM2000, LieblingPRD2002, LieblingCQG2004} while using time-marching, finite-difference based methods. Adaptive, high resolution methods are needed in general in order to properly resolve the transition region and find self-similar solutions. While this has generally been done with adaptive mesh refinement box-in-box style finite difference methods in the past, finite element methods (FEM) including both continuous and discontinuous Galerkin approaches are appearing in numerical relativity studies as an effective alternative in studying astrophysical collapse~\cite{diener}. Similarly, this study re-examines collapse and dispersion in the semilinear wave equation using a novel continuous space-time Galerkin method to explore the benefits of FEM approaches in studying collapse.

Finite element methods (FEM) are way to form and solve the discrete versions of certain differential equations. Instead of finding a way to approximate the derivatives, these methods require the reformulation of the differential equation into a variational problem with a function space where the solution can exist. Then, the domain is divided into a finite number of elements, and a piecewise basis function is applied to each node on the domain to create discrete function space. The original problem is thus converted into a discrete problem for which the solution can be numerically approximated. While finite difference methods are very useful and sometimes very easy to implement, the FEM's main advantages lie in their ability to handle complicated domain and boundary geometries. Simplicity of the domain cannot be guaranteed in practical applications, and for this reason, we choose to implement the FEM in this research. 

Adaptive space-time FEM approaches for solving systems of nonlinear equations present numerous numerical challenges while also enabling more asynchronous operation and the extraction of more parallelism when utilizing high performance computing resources. For certain formulations of partial differential equations (PDE), time-parallel preconditioners may also be successfully applied in space-time finite element simulations with substantial improvements in scalability. A space-time FEM increases concurrency on distributed memory machines by permitting time-decomposition in addition to spatial-decomposition. This paper applies a space-time FEM previously introduced~\cite{AndersonJCP2007} to the semilinear wave equation with nonlinear term $p=7$. The semilinear wave equation with $p=7$ has been shown critical behavior near singularity formation similar to that found in black hole critical behavior~\cite{LieblingPRD2008, BizonNL2004, BizonNL2007} making it a good test problem for exploring this space-time FEM. Because of the many length and time scales involved in solving the semilinear wave equation, this also presents an opportunity to examine space-time finite elements with a nonuniform space-time mesh as opposed to more conventional adaptive mesh refinement techniques. 

In addition to the parallel computing benefits resulting from increased concurrency, space-time finite element approaches have other advantages for numerical simulations. The space-time FEM explored here is a fully-implicit method, and it can use time-varying computational domains, higher order approaches, and unstructured meshes. It has been extended for 2+1 and 3+1 dimensions with efficiency and accuracy~\cite{AndersonJCP2007}. The major disadvantage of the approach is the significant memory overhead requirement that the entire space-time problem must now fit in memory all at once. However, time parallel approaches with non-uniform meshes have not been generally used in the scientific computing community in the past and the increased concurrency benefits may outweigh the significant memory overhead challenges and other barriers to implementation.

The paper is organized as follows: in Section 2, the work related to this is presented; in Section 3, a background on the semilinear wave equation is provided; in Section 4, the numerical approach is provided, including the 1+1 and 2+1 space-time finite element discretization, a time parallelizable preconditioner based on the additive Schwarz method, the boundary conditions, and mesh refinement; Section 5 presents results while Section 6 contains the conclusions.

\section{Related Work}

The threshold of singularity formation in the semilinear wave equation for different values of $p$ was explored in depth by Liebling~\cite{LieblingPRD2008}using conventional finite difference techniques and Crank-Nicolson integration. The existence of self-similar solutions and singularity formations in the semilinear wave equation with a focusing nonlinearity was examined in~\cite{BizonNL2004, BizonNL2007}. The space-time finite element method used for this work was introduced in~\cite{AndersonJCP2007} where the space-time approach is implemented along with time decomposition methods for a nonhomogeneous wave equation. 
Similarly, ~\cite{LimSURIO2013} implements the Klein-Gordon equation in 1(space)+1(time) dimensions also based on the numerical method in~\cite{AndersonJCP2007}. ~\cite{DonMC1996} presents continuous finite elements in time and space simultaneously to solve the wave equation. While Continuous Galerkin approaches with space-time FEM can be found in several engineering examples, including ~\cite{CsikIJNMF2002, KimJSME2001, IdeCM2001, GuddatiISS1999, KitIJNME1997}, they have not yet been applied to the semilinear wave equation and investigating critical behavior apart from this work. Some efforts at discontinuous Garlerkin space-time finite element approaches have also been investigated~\cite{RheJCP2008, AbeIJSS2011} though outside the context of critical behavior.

The semilinear wave equation has been studied in several ways. ~\cite{YangAMC2006} used a finite difference discretization and a shooting method as a solving technique. ~\cite{LieblingPRD2013} explored the semilinear equations in spherical symmetric AdS space to show nonlinear collapse. ~\cite{MikPRE2006} focuses on variational solutions of the semilinear wave equation with space-time fractional Brownian noise. ~\cite{OliNM2004} presents the proof of a standard second order finite difference uniform space discretization of the semilinear wave equation with periodic boundary conditions. 
In general, the semilinear wave equation has proven to be a useful model for exploring the robustness of a numerical method in many past studies.

\section{Model Problem}

\subsection{The Semilinear Wave Equation}
\begin{equation}\label{eqn:semil}
\left(\frac{1}{c^2}\frac{\partial^2}{\partial t^2}-\nabla^2+\frac{m^2 c^2}{\hbar^2}\right)\Psi(\mathbf x,t) = \Psi^p(\mathbf x,t)
\end{equation}

The semilinear wave equation is given in Eqn~\ref{eqn:semil}. The nonlinear term, $\Psi^p(\mathbf x,t)$, determines the singularity formation behavior. 
To preserve the symmetry ($\Psi \rightarrow -\Psi$), odd powers of $p$ are used; previous work has examined the cases with $p=3,5,7$~\cite{LieblingPRD2008}. The semilinear wave equation represents one of the simplest nonlinear generalizations of the wave equation, and it shows interesting behavior at the threshold of singularity formation. For the $p = 3$ case, the threshold could not be found while for the $p = 5$ case, a critical solution is observed, and self-similar solutions appear roughly, but there is no non-trivial self-similar solution. 
Instead, the solutions approach scale evolving to static solutions. For the $p = 7$ case, a critical solution is found which approaches the self-similar solution in spherical symmetry. Consequently the $p=7$ case is selected for further investigation using the space-time finite element method.

Reference ~\cite{BizonNL2007} shows the existence of self-similar solutions of the semilinear wave equation. According to~\cite{BizonNL2004}, self-similar solutions can be found using time scale transformations:
\begin{equation}\label{eqn:scalet}
\Psi(r,t)=(T-t)^{-\xi} U(\rho)
\end{equation}
where
\begin{equation}
\xi=\frac{2}{p-1} \indent \rho=\frac{r}{T-t}
\label{eqn:transform}
\end{equation}
where $p$ is the nonlinear power term of this equation, and $T$ is a certain collapse time. 
As shown in ~\cite{BizonNL2004}, self-similar solution can be obtained by rewriting the semilinear wave equation.
Substituting Eqn.~\ref{eqn:scalet} into Eqn.~\ref{eqn:semil}, the ordinary differential equation shown in Eqn.~\ref{eqn:shoot} for finding the self-similar solution results.
\begin{equation}
(1-\rho^2)U''(\rho)+\left(\frac{2}{\rho}-(2+2\alpha)\rho \right)U'(\rho)-\alpha(\alpha+2)U(\rho) + U^p(\rho)=0
\label{eqn:shoot}
\end{equation}

Reference~\cite{BizonNL2004} solves Eqn.~\ref{eqn:shoot} using the shooting technique. 
Applying this technique, Biz\'{o}n and Maison proved the existence for a countable set of parameters which determine explicit self-similar solution $U_n(\rho)$. 
This allows the numerical results presented here to be compared with solutions $U_1(\rho)$ provided from the previous study~\cite{BizonNL2004} for the $p=7$ case with the same time scale transformations given in Eqn.~\ref{eqn:transform}.

\section{Numerical Approaches}

\subsection{Space-Time Finite Element Method}
A space-time finite element method using continuous approximation functions in both space and time is used to explore numerical singularity formation. The discretization of the semilinear wave equation in this paper is an extension 
of the discretization of the nonhomogeneous wave equation presented in~\cite{AndersonJCP2007}. 

The semilinear wave equation with natural units in 1+1 dimensions is
re-written in Eqn.~\ref{eqn:semil2}:

\begin {equation} \label{eqn:semil2}
\frac{\partial^2 \Psi}{\partial t^2}-\frac{\partial^2 \Psi}{\partial x^2}+\Psi=\Psi^7
\end {equation}

Space and time are discretized together for the entire domain 
using a finite element space which does not discriminate between space and 
time basis functions. Iterative solution methods in conjunction 
with a time decomposition preconditioner are employed for the solution.
Introducing two new variables, $u=\Psi$ and 
$v=\frac{\partial \Psi}{\partial t}$,
the system is re-written into first-order in time form, Eqn.~\ref{eqn:first_order}:

\begin {eqnarray} 
-\frac{\partial^2 u}{\partial x^2}+\frac{\partial v}{\partial t} +u = u^7 \nonumber \\
-\frac{\partial u}{\partial t}+v=0 \label{eqn:first_order}
\end {eqnarray}

The finite element space is the space of piecewise polynomial functions $\phi : \Omega \times (0,T] \to \mathfrak{R}$ where $\Omega \equiv (0,d]$, and $d$ is a physical space size of the system.

In spherical symmetry, this equation is re-written with respect to one spherical coordinate, given in Eqns.~\ref{eqn:Klein-Gordon2}--\ref{eqn:Klein-Gordon3}.

\begin{equation}\label{eqn:Klein-Gordon2}
\frac{\partial^2 \Psi}{\partial t^2} - \frac{1}{r^2} \frac{\partial}{\partial r}\left( r^2 \frac{\partial \Psi}{\partial r}\right)+\Psi = \Psi^7
\end{equation}

\begin{equation}\label{eqn:Klein-Gordon3}
\frac{\partial^2 \Psi}{\partial t^2} - \frac{\partial^2 \Psi}{\partial r^2} -\frac{2}{r}\frac{\partial \Psi}{\partial r}+\Psi = \Psi^7
\end{equation}

Applying the auxiliary variables $u=\Psi$ and $v=\frac{\partial \Psi}{\partial t}$ to the system results in Eqns.~\ref{eqn:aux0}--\ref{eqn:aux}.

\begin {eqnarray}
\label{eqn:aux0}
-\frac{\partial^2 u}{\partial r^2}-\frac{2}{r} \frac{\partial u}{\partial r}+\frac{\partial v}{\partial t} + u = u^7 \\
-\frac{\partial u}{\partial t}+v=0
\label{eqn:aux}
\end {eqnarray}

The weak form of these equations is shown in Eqns.~\ref{eqn:weak1}--\ref{eqn:weak2}:
\begin{eqnarray}
\label{eqn:weak1}
K(u,v,\phi) = \int_{\Omega} \left(-\frac{\partial u}{\partial r} \frac{\partial \phi}{\partial r} -\frac{2}{r}\frac{\partial u}{\partial r}\phi+\frac{\partial v}{\partial t} \phi +u\phi- u^7\phi \right) ds = 0 \\
\label{eqn:weak2}
G(u,v,\phi) = \int_{\Omega} \left(-\frac{\partial u}{\partial t} \phi + v \phi \right) ds = 0 
\end{eqnarray}

For discretization, a single square element with rectangular basis functions is considered. The element stiffness matrix is used to assemble the stiffness matrix for the entire domain. 
In the event there is no mesh refinement, all 
the elements in the square of $\Omega$ are the same size and $A_j$ 
can be utilized for all elements by only having to adjust for the orientation 
of the square. In mesh refinement case, the element matrices $A_j$ 
are assembled with respect to different basis functions within the 
different mesh refined regions.

The basis functions for $1+1$ case are determined explicitly at the nodes:~\cite{AndersonJCP2007, LimSURIO2013}

\begin {eqnarray}
\phi_A = \frac{1}{h^2} xt-\frac{1}{h} x- \frac{1}{h} t +1 \\
\phi_B = -\frac{1}{h^2} xt +\frac{1}{h} x \\
\phi_C = \frac{1}{h^2} xt \\
\phi_ D= -\frac{1}{h^2} xt + \frac{1}{h}t 
\end {eqnarray}
Using those basis functions the element stiffness matrix, $A_j$, is assembled. The element stiffness matrix for this singular element is represented as:
\begin{displaymath}
\mathbf{A_j}=
\left[ \begin{array}{cccc}
a_j(\phi_A,\phi_A) & a_j(\phi_A,\phi_B) & a_j(\phi_A,\phi_C) & a_j(\phi_A,\phi_D) \\ 
a_j(\phi_B,\phi_A) & a_j(\phi_B,\phi_B) & a_j(\phi_B,\phi_C) & a_j(\phi_B,\phi_D) \\
a_j(\phi_C,\phi_A) & a_j(\phi_C,\phi_B) & a_j(\phi_C,\phi_C) & a_j(\phi_C,\phi_D) \\
a_j(\phi_D,\phi_A) & a_j(\phi_D,\phi_B) & a_j(\phi_D,\phi_C) & a_j(\phi_D,\phi_D) 
\end{array} \right]
\end{displaymath}
where each element of the matrix $A_j$ $(a_j(\phi_\alpha, \phi_\beta)$, and $\alpha, \beta = A, B, C, D)$ can be calculated by integration of the weak form in Eqns.~\ref{eqn:weak1}--\ref{eqn:weak2}.

For the $2+1$ case, the finite element space is the space of piecewise polynomial functions $\phi : \Omega \times \Omega \times (0,T] \to \mathfrak{R}$ where $\Omega \equiv (0,d]$, and $d$ is a physical space size of the system which is same as the $1+1$ case. Then, the basis functions are:
\begin{eqnarray}
\phi_1 = \frac{1}{h}x-\frac{1}{h^2}xy-\frac{1}{h^2}xt+\frac{1}{h^3}xyt\\ 
\phi_2 = \frac{1}{h^2}xy-\frac{1}{h^3}xyt \\
\phi_3 = \frac{1}{h}y-\frac{1}{h^2}xy-\frac{1}{h^2}yt+\frac{1}{h^3}xyt\\ 
\phi_4 =1- \frac{1}{h}x-\frac{1}{h}y-\frac{1}{h}t+\frac{1}{h^2}xy+\frac{1}{h^2}xt+\frac{1}{h^2}yt-\frac{1}{h^3}xyt\\ 
\phi_5 = \frac{1}{h^2}xt-\frac{1}{h^3}xyt \\
\phi_6 = \frac{1}{h^3}xyt \\
\phi_7 = \frac{1}{h^2}yt-\frac{1}{h^3}xyt \\
\phi_8 = \frac{1}{h}t-\frac{1}{h^2}xt-\frac{1}{h^2}yt+\frac{1}{h^3}xyt
\end{eqnarray}
Using above fnctions, the element stiffness matrix for a $2+1$ case is represented as:
\begin{displaymath}
\mathbf{A_j}=
\left[ \begin{array}{ccc}
a(\phi_1, \phi_1) & \cdots & a(\phi_1, \phi_8)\\
\\
\vdots & \ddots & \vdots \\
\\
a(\phi_1, \phi_8) & \cdots & a(\phi_8, \phi_8) \\
\end{array} \right]
\end{displaymath}
where each element of the matrix $A_j$ $(a_j(\phi_\alpha, \phi_\beta)$, and $\alpha, \beta = 1, \cdots , 8)$ also can be calculated by integration of the weak form in Eqns.~\ref{eqn:weak1}--\ref{eqn:weak2}.

The stiffness matrix $A$ is an $N \times N$ matrix, where $N$ is the total number of nodes in the domain. In the unit square example, $N = W^2$ where $W$ is the number of collocation points in the mesh.\\
The stiffness matrix $A$ is defined as:
\begin{displaymath}
A=
\left[ \begin{array}{ccc}
a(\phi_1, \phi_1) & \cdots & a(\phi_1, \phi_{W^2})\\
\\
\vdots & \ddots & \vdots \\
\\
a(\phi_{W^2}, \phi_1) & \cdots & a(\phi_{W^2}, \phi_{W^2}) \\
\end{array} \right]
\end{displaymath}
The stiffness matrix component in row $i$ and column $k$ as $a_{ik}$.
The stiffness matrix component in row $i$ and column $k$, 
$a_{ik} = a(\phi_i, \phi_k)$ can be calculated by adding the effects 
of all the square elements: 
\begin{equation*}
a_{ik} =\sum \limits_{T_j \in \Omega} a_j (\phi_i, \phi_k). 
\end{equation*}

Newton's method is used as a nonlinear solver. The Jacobian matrix 
is constructed with the finite element space.

Then, a linear system of equations $Ax=b$ is constructed. Because of the large size of the discretized problems, iterative numerical methods based on Krylov subspace (KSP) methods are used. A proper preconditioner is also needed to help convergence and increase the speed of convergence. The additive Schwarz preconditioner (ASM)~\cite{saaditer, Toselli:2004:DDM, Smith:1996:DPM} is employed for the numerical simulation. Applying the ASM to space-time finite elements using a time decomposition method is an important aspect of the
solution of the space-time FEM approach used here.

\subsection{Additive Schwarz Method}

Domain Decomposition methods (DD) solve a boundary value problem by splitting it into smaller boundary value problems on subdomains and iterating to coordinate the solution between adjacent subdomains. The problems in the subdomains are independent, which makes domain decomposition methods suitable for parallel computing. Domain decomposition methods are typically used as preconditioners for Krylov space iterative methods, such as the conjugate gradient method or GMRES. Domain decomposition methods show large potential for a parallelization of finite element methods in general, and serve as a basis for distributed, parallel computations.

The time decomposition methods in this paper are a variant of the time additive Schwarz method (ASM). 
For a domain $\Omega = \cup_i \Omega_i$, the ASM can be written 
as Eqn.~\ref{eqn:asm}:
\begin{equation}
x^{n+1} = x^n + \sum_i B_i(f-Ax^n) 
\label{eqn:asm}
\end{equation}
where $x$ is the solution vector of the linear system $Ax=b$, $A$ is the matrix representation of the system. And, let $R_i$ is the restriction to $\Omega_i$ and let $A_{\Omega_i} = R_i A R_i^T$ which is restricted operator for the interior grid points in $\Omega_i$. Then, $B_i=R_i^T A^{-1}_{\Omega_i} R_i$.
The additive Schwarz method may also be viewed as a generalization of block Jacobi methods~\cite{saaditer, Toselli:2004:DDM, Smith:1996:DPM}.

\subsection{Implementation Components}
\begin{figure}[h]
\centering
\captionsetup{justification=RaggedRight}
\includegraphics[width=0.6\textwidth]{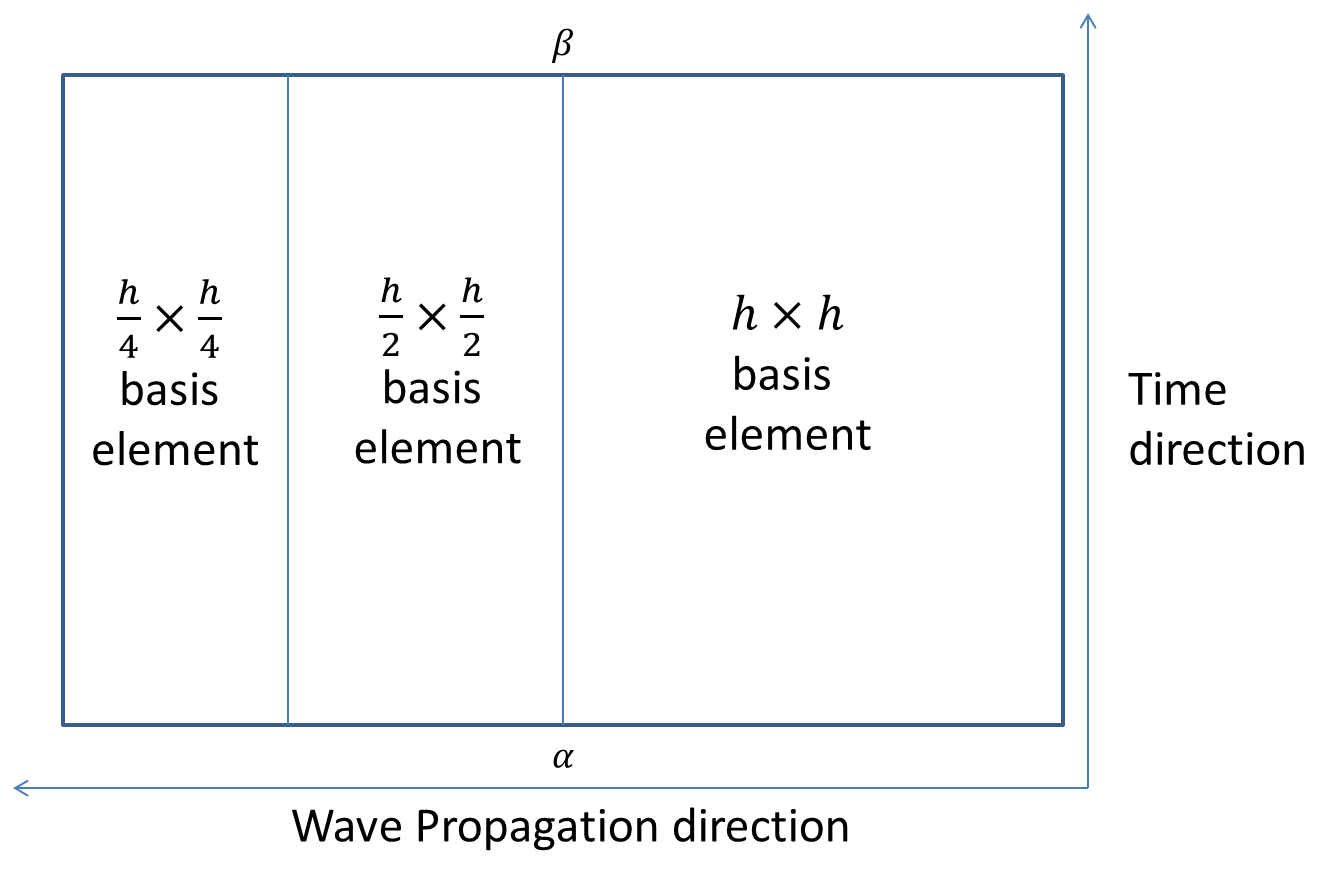} 
\caption{The general scheme of the domain with mesh refinement. Robin boundary conditions are applied at $r=0$ and $r=r'$. Several combinations of the two parameters, $\alpha$ and $\beta$, are explored for use at the different boundaries.} 
\label{fig:figdo}
\end{figure}
In this problem, the semilinear wave equation is solved in spherical symmetry, resulting in a switch to spherical
coordinates as well as a difficulty with the $r=0$ boundary. 
The Robin boundary condition is used for the boundaries
with parameters $\alpha$ for the $r=0$ boundary and $\beta$ for the outer boundary:
\begin{eqnarray}
u+\alpha \frac{\partial u}{\partial n} = 0 \indent u+ \beta \frac{\partial u}{\partial n} = 0
\end{eqnarray}
Figure~\ref{fig:figdo} shows the general scheme of this problem.
The appropriate values of $\alpha$ and $\beta$ are found empirically.
Also, mesh refinement is used in order to improve computational efficiency.
Figure~\ref{fig:figdo} shows the strategy of mesh refinement employed here. Mesh refinement adds resolution to the mesh
when and when it is needed. As solution behavior near $r = 0$ shows the largest gradients and error, that region is discretized using a finer mesh than in other regions. 
The generalized minimal residual method (GMRES)~\cite{SaadSIAM1986} is used to solve unsymmetrical system of the discreitzed matrix $A$. And, GMRES is used with a preconditioning method~\cite{saaditer, Toselli:2004:DDM, Smith:1996:DPM}  in order to speed up convergence.

\section{Numerical Results}
\label{sec:results}

\subsection{Critical behavior}

PETSc~\cite{petsc-web-page} is used for implementing the time-additive Schwarz precondtioner and
for using GMRES. After performing a large
number of simulations exploring critical behavior, the results are compared with previous 
research~\cite{LieblingPRD2008, BizonNL2004, BizonNL2007}. The existence of self-similar solutions in those
results are also explored.

A Gaussian pulse is used for initial data
\begin{equation}
\Psi(r,t)_{ini}=A e^{-(r-R)^2}
\label{eqn:gauss}
\end{equation}
where $A$ is the amplitude, and $R$ is the initial radius of the pulse. We observe critical behavior with $A=0.1720$ as an initial amplitude value, $\alpha = 0.01$ and $\beta=-100$ for Robin boundary values, and $R=8$ for an initial data radius. 

\begin{figure}[h!]
\centering
\captionsetup{justification=RaggedRight}
\includegraphics[width=0.65\textwidth]{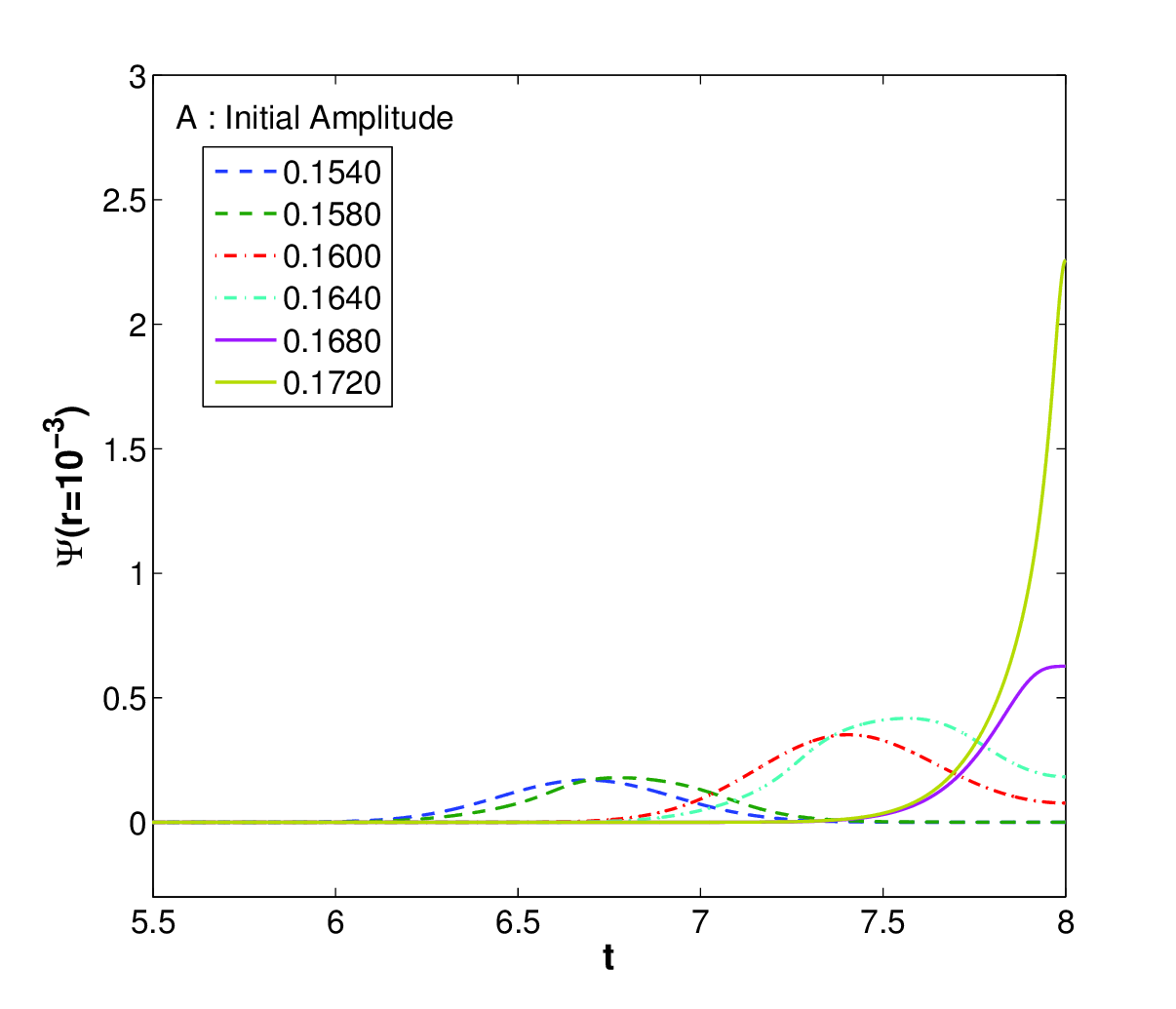} 
\caption{Solution plots in the time domain with different initial amplitude values $A$. Three different amplitude value families are created, defined as: $F_1= 0.1540$ and $0.1580$, $F_2=0.1600$ and $0.1640$, and $F_3 = 0.1680$ and $0.1720$. Each family has a small difference in amplitude values (at the 3 significant digits). In order to investigate the effects of initial data for the singularity formations, the value of $\Psi$ is plotted until the wave reaches the collapse time of $T \simeq8$ around the $r=0$ region ($r \simeq 10^{-3}$). The initial Gaussian pulse is the same as in Eqn.~\ref{eqn:gauss} for the different families of initial amplitude values $A$.} 
\label{fig:soltime}
\end{figure}

Figure \ref{fig:soltime} shows value of the $\Psi$ plotted as a function of time. For specific tests, amplitude ranges are selected near $A=0.1720$, and three families are created for a heuristic test. Solutions of $F_1$ and $F_2$ disperse before the collapse time, and the properties of the results are very similar. However, solutions of $F_3$ show different properties. $A=0.1680$ also does not show a critical evolution, but $A=0.1720$ presents a critical evolution near $r=0$. 

The results indicate that the solutions of the semilinear wave equation with $p=7$ case disperse for some initial data and blow up for some other initial data even though those initial data are not very different. The determination of the threshold of singularity formation and the corresponding dynamics is a great interest for future studies.

\begin{figure}[h!]
\centering
\captionsetup{justification=RaggedRight}
\includegraphics[width=0.65\textwidth]{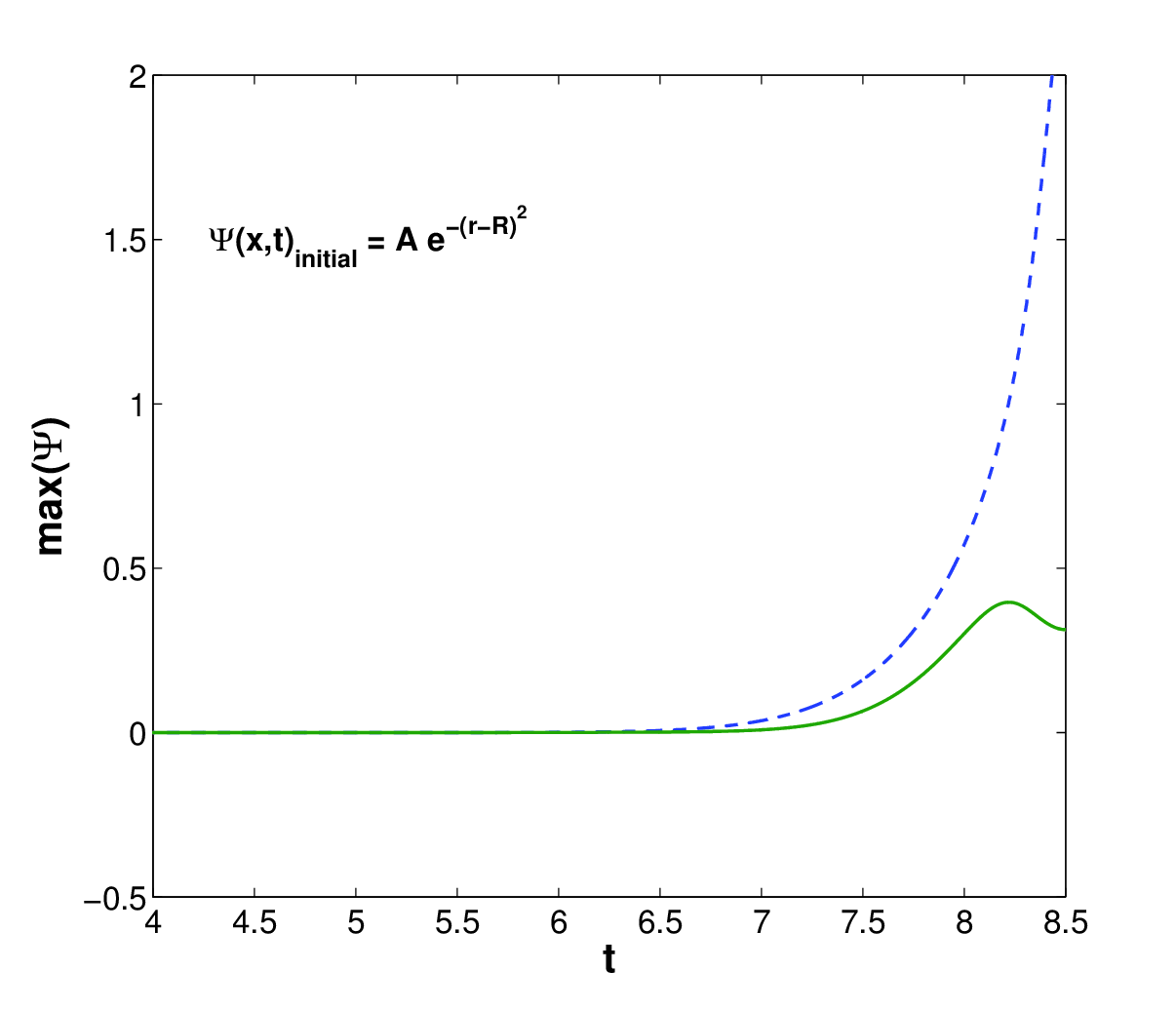} 
\caption{Plots of the near critical evolution region $r=0$. Simulations are performed using the same initial 
data as in the previous Figure~\ref{fig:soltime} but with slightly different initial amplitude values. The maximum of the $\Psi$ field is 
plotted versus time. The green solid line shows a slightly subcritical evolution, and the blue dotted line shows a slightly supercritical evolution that is close to critical evolution. In this study, critical collapse occurs at $T \simeq 8$.} 
\label{fig:solnct}
\end{figure}

For more specific views near critical behavior, Figure~\ref{fig:solnct} shows two near critical evolutions which occur near the collapse time $T \simeq 8$.
The supercritical solution (blue dotted) and subcritical solution (green solid) for the $p=7$ case can be observed to change with the initial amplitude values.

\begin{figure}[h]
\centering
\captionsetup{justification=RaggedRight}
\subfloat[Plot for non-critical evolution]{
\includegraphics[width=0.45\textwidth]{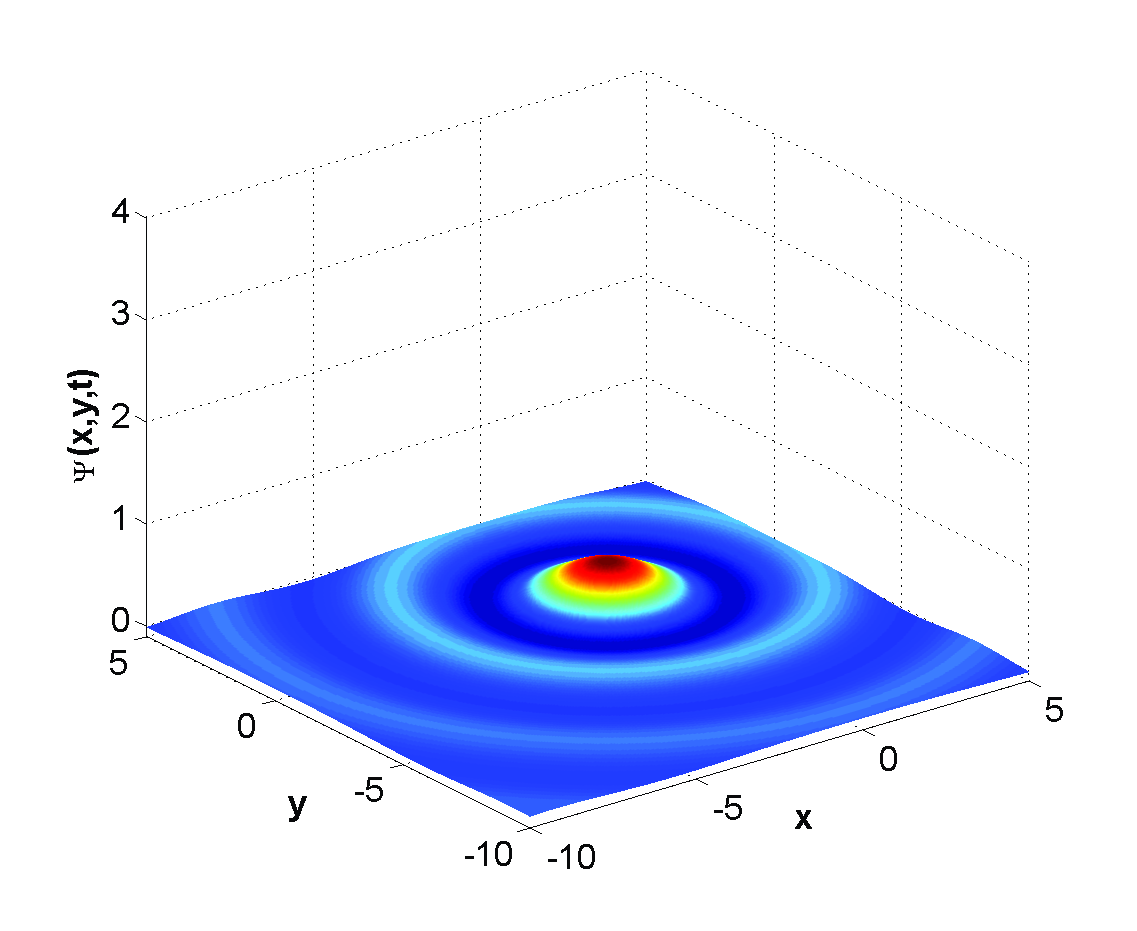} }
\subfloat[Plot for critical evolution]{
\includegraphics[width=0.45\textwidth]{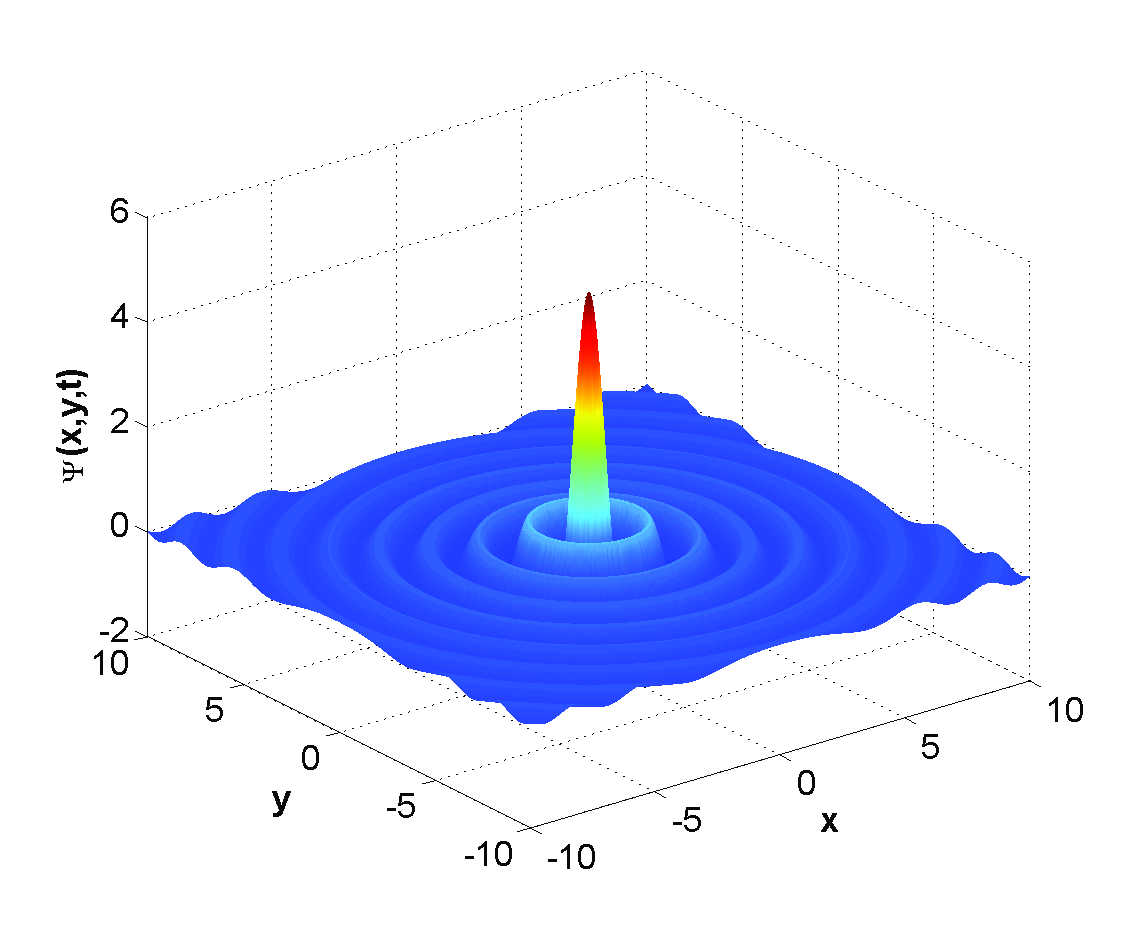} }\\
\subfloat[Solution plots for one spatial dimension with different initial amplitude values at critical collapse time T $\simeq$ 8]{
\includegraphics[width=0.65\textwidth]{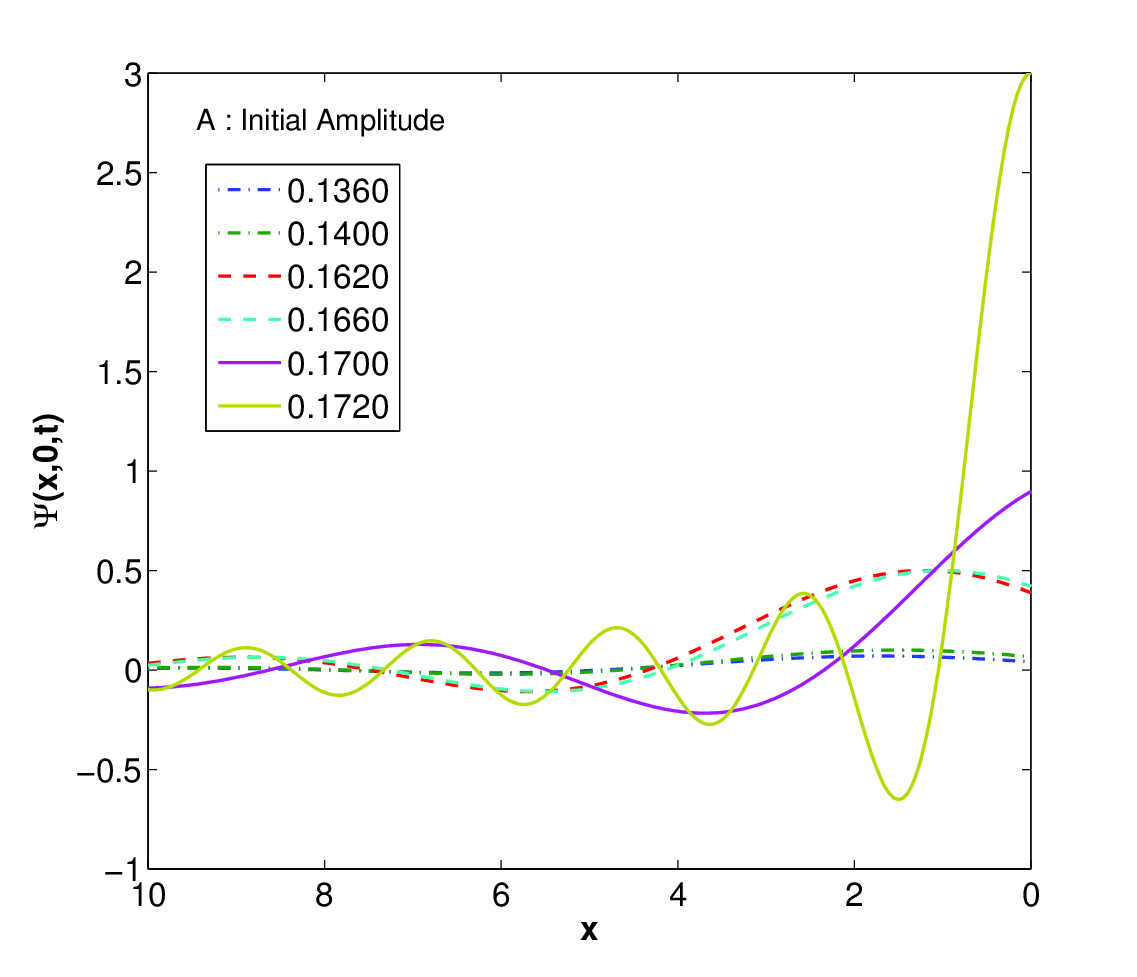} }
\caption{Solution plots of the semilinear wave equation for the $2+1$ case at the collapse time $T \simeq8$. The initial data is a 2D Gaussian pulse. In subfigures~\ref{fig:3d1}(a) and (b), two different amplitude values are chosen to investigate critical behavior. Based on the information in previous subfigures, subfigure~\ref{fig:3d1}(c) shows value of the $\Psi$ plotted as a function of one spatial dimension ($x$) for heuristic tests. Three different amplitude value $A$ families are created, defined as: $F_1= 0.1360$ and $0.1400$, $F_2=0.1620$ and $0.1660$, and $F_3 = 0.1720$ and $0.1720$. Each family has a small difference in amplitude values. The value of $\Psi$ is plotted at the collapse time of $T \simeq8$ around the $r=0$ region ($r \simeq 10^{-3}$).}
\label{fig:3d1}
\end{figure}


Figure~\ref{fig:3d1} shows plots for the $2+1$ case. In subfigure~\ref{fig:3d1}(a), plot does not show critical behavior, but plot shows critical evolution as a spherical peak in subfigure~\ref{fig:3d1}(b). For specific tests, amplitude ranges are selected near $A=0.1720$, and three families are plotted for a heuristic test. Solutions of $F_1$ and $F_2$ disperse before the collapse time, and the properties of the results are very similar. However, solutions of $F_3$ show different properties. $A=0.1700$ also does not show a critical evolution, but $A=0.1720$ presents a critical evolution near $r=0$ at $T \simeq 8$. This result is similar to the $1+1$ 
case (Figure~\ref{fig:soltime}). The success of the space-time FEM approach in
the $1+1$ and $2+1$ cases suggests the feasibility to extend the work to
higher dimensions such as $3+1$ using tesseractic elements.

\begin{figure}[h!]
\centering
\captionsetup{justification=RaggedRight}
\includegraphics[width=0.65\textwidth]{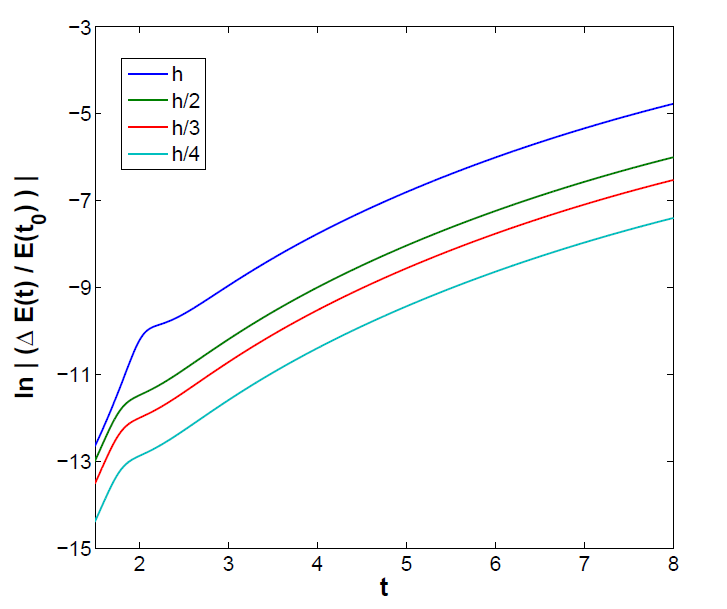} 
\caption{Enery loss plots for the $p=7$ case based on different resolutions. The $h$ is a space resolution which can be obtained by dividing the \textit{physical domain size by the number of collocation points}, or $15/5000 = 0.003$. The value of $h=0.003$ is base resolution which is chosen arbitrary. The energy loss ($\Delta E(t) \equiv E(t) -E(t_0)$) is plotted until the initial data reaches the critical evolution time $t_c=T \simeq 8$. As the resolution increases, energy loss decreases. The results of the convergence test is $(\| \Delta E_{h/4} - \Delta E_{h/2} \|_2) / (\| \Delta E_{h/2} - \Delta E_{h} \|_2)=4.133$. This result indicates that the order of energy loss self convergence is second order.} 
\label{fig:solener}
\end{figure}

Figure~\ref{fig:solener} shows energy loss between initial time and critical evolution time. Ideally, the energy loss ($ | \Delta E(t)|$) should be zero. A convergence study to test this was undertaken where the same initial conditions were used but with different resolutions. The self convergence test is $(\| \Delta E_{h/4} - \Delta E_{h/2} \|_2) / (\| \Delta E_{h/2} - \Delta E_{h} \|_2)=4.133$ which indicates second order convergence.
\subsection{Self-Similarity}

\begin{figure}[h!]
\centering
\captionsetup{justification=RaggedRight}
\subfloat[Demonstration of self-similarity for $1+1$ case]{
\includegraphics[width=0.65\textwidth]{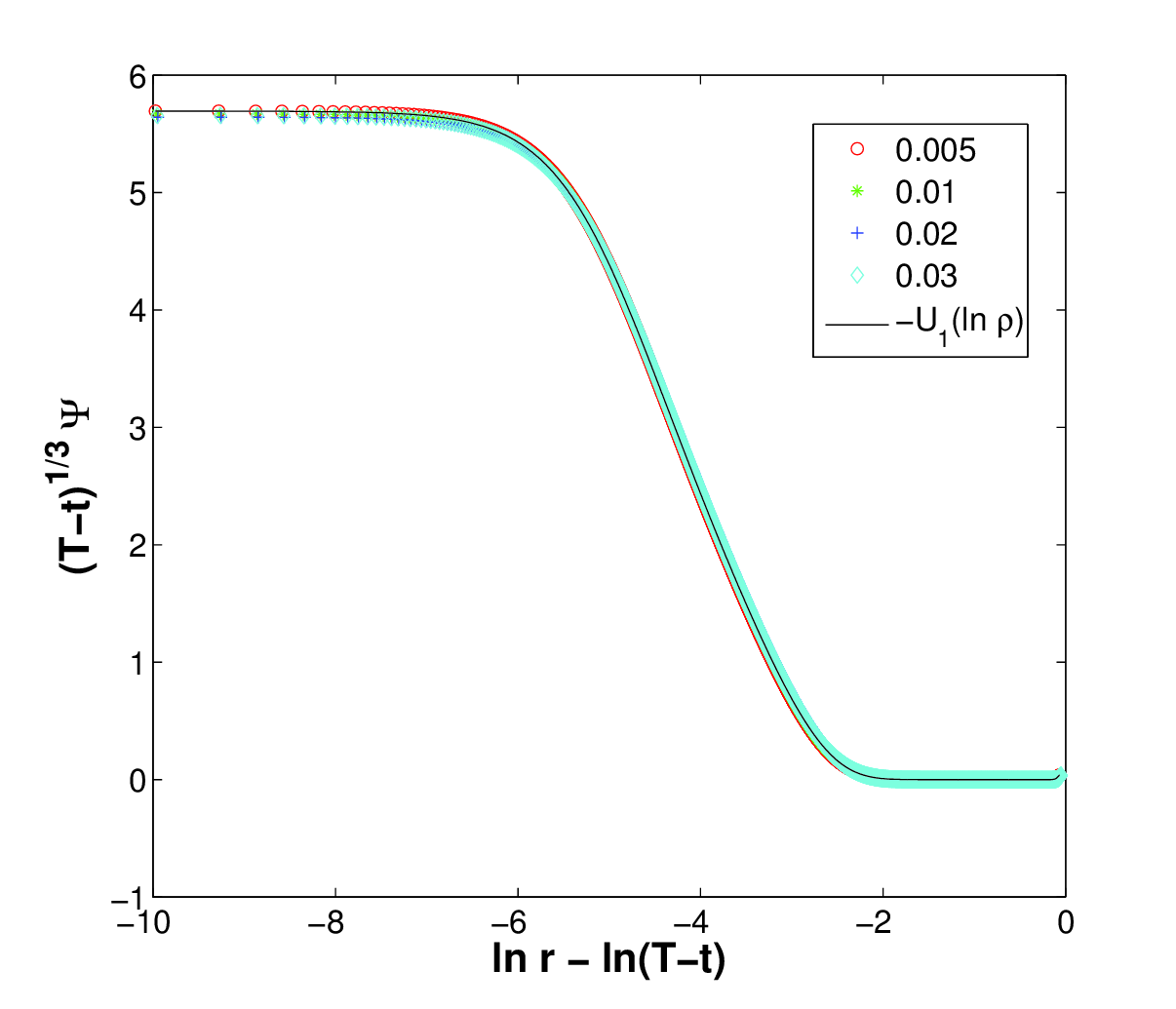} }\\
\subfloat[Demonstration of self-similarity for $2+1$ case]{
\includegraphics[width=0.65\textwidth]{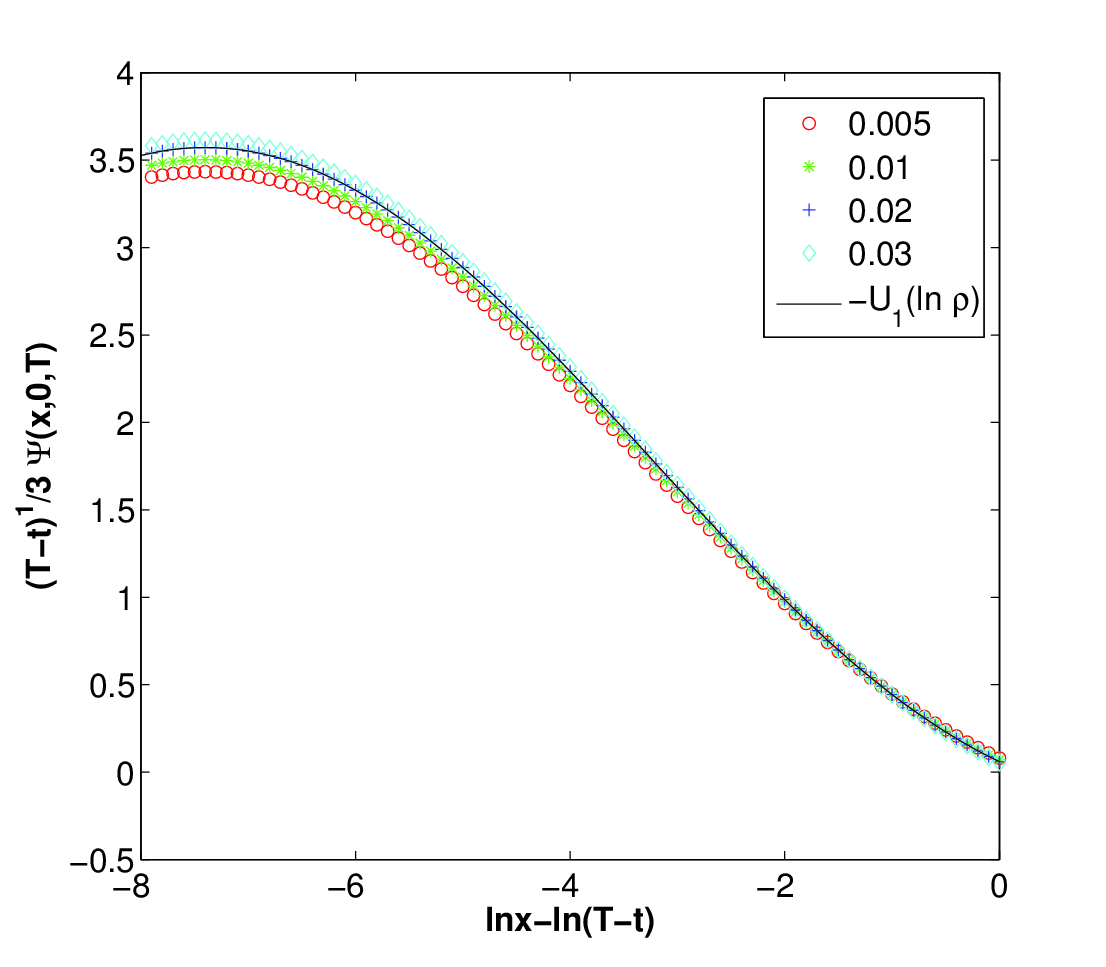} }
\caption{Determination of self-similar solutions. The same initial data values as illustrated above (R=8, A=0.1720 with Robin boundary values: $\alpha=0.01$, $\beta=-100$) are chosen for the both $1+1$ and $2+1$ tests. The collapse time is $T \simeq 8$. Different time scale transformations are performed that based on that information. Define $T-t=\epsilon$ as a small parameter for time differences where $T$ is the collapse time for this system. The simulations are performed with $\epsilon= 0.005, 0.01, 0.02$, and $0.03$. This figure also shows the compared results with a solution of the ordinary differential equation (23). The black solid line is the explicit self-similar solution $U_1(\rho)$ from reference~\cite{BizonNL2004}. The log scale transformation for $U_1(\rho)$ is applied to obtain the compared results. The graphs show the results $-U_1(\ln \rho)$ with the numerical data time scale transformed 
based on $T-t=\epsilon$}
\label{fig:ssode}
\end{figure}

Self-similar solutions are often found at the threshold of critical behavior.
We consider Eqn.~\ref{eqn:shoot} inside the critical behvior region $(t=T, r=0)$, that is in the interval $0 \leq \rho \leq 1$. The test problem is the $p=7$ case, and paper~\cite{BizonNL2004} shows solutions of the ODE by the shooting technique. Applying this technique, Biz\'{o}n and Maison proved the existence for a countable set of parameters $b_n (n = 0,1,...)$ which determine explicit self-similar solution $U_n(\rho)$ for $p=3$ and all odd $p \geq 7$.
Therefore, our numerical results can be compared with ODE solutions $U_1(\rho)$ from the previous study~\cite{BizonNL2004} for the $p=7$ case within the same time scale transformations in Figure~\ref{fig:ssode}.

Figure~\ref{fig:ssode} shows the compared results with a solution of the ODE. Four different solutions transformed using 
Eqn.~\ref{eqn:scalet} are shown. The different time scaled transformation solutions coincide showing self-similarity for both $1+1$ and $2+1$ cases. The time scale transformed solutions coincide with the $-U_1(\ln \rho)$, and this indicates that the solutions present self-similarity. 

\clearpage

\subsection{Performance Tests}

Performance results from simulations using the space-time FEM both with and without
time decomposition are presented in this section. A self-similar solution
resulting from the space-time FEM simulations using mesh
refinement is also presented. Performance results were found using a cluster of 
Intel Xeon E5-2690 2.90 Ghz processors.

\begin{figure}[h]
\centering
\captionsetup{justification=RaggedRight}
\includegraphics[width=0.65\textwidth]{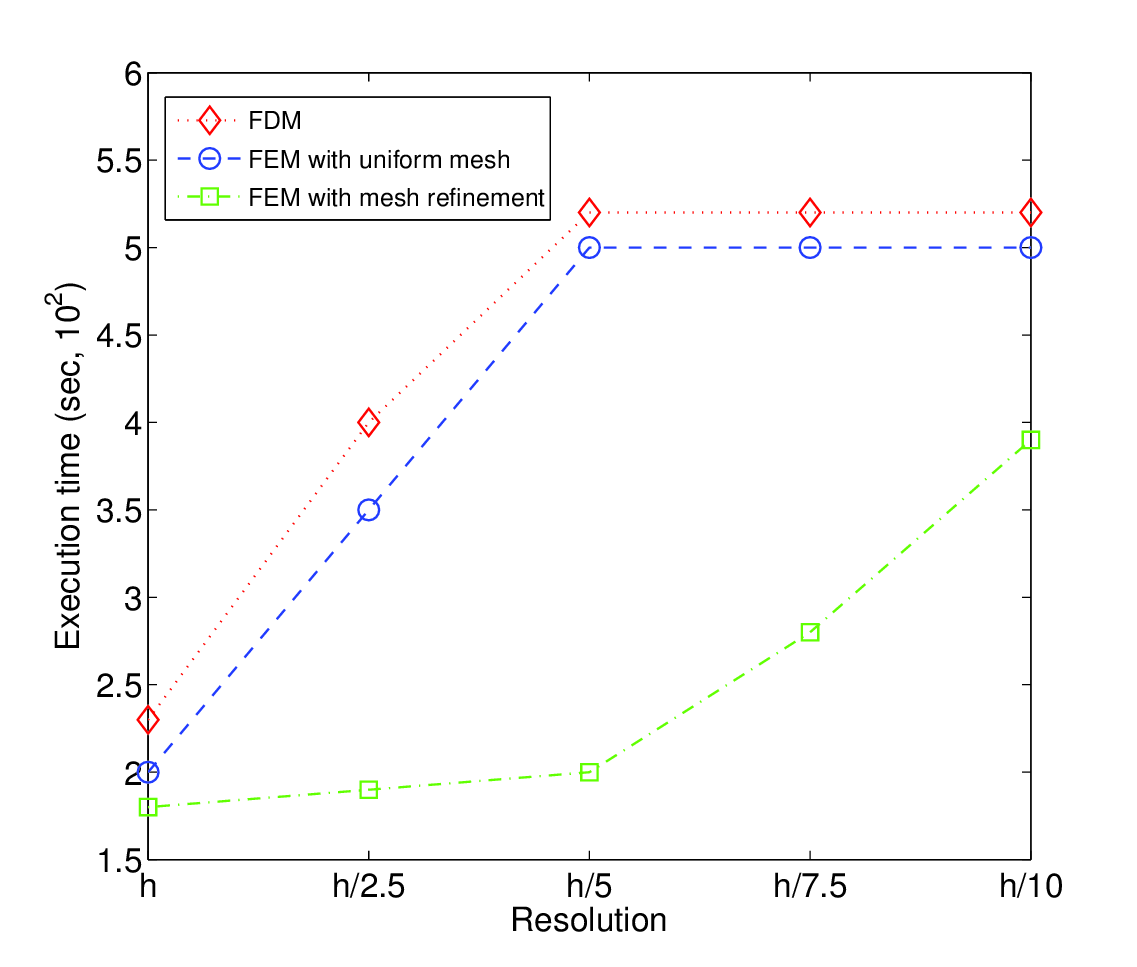} 
\caption{Execution time measured for different numerical methods employed
to solve the semilinear wave equation: the finite difference method, 
the space-time FEM with a uniform mesh, and the space-time FEM 
with mesh refinement. The same initial conditions as in Figure~\ref{fig:ssode} with different resolutions are applied to the tests. The additive Schwarz preconditioner, which accelerates the performance even further, was not used for these tests but
is reported in Table~\ref{tbl:compn}. The resolution comparisons are based on the finest grid region (near $r=0$). The $h$ is the space resolution which is calculated as in Figure~\ref{fig:solener}, yielding $15/5000=0.003$.}
\label{fig:compm}
\end{figure}

In Figure~\ref{fig:compm} and Table~\ref{tbl:convt}, comparisons 
of performance and final solution value when solving the 
semilinear wave equation are provided. These results compare using
the finite difference method, the space-time finite element method 
with a uniform mesh, and the space-time finite element with mesh 
refinement to each solve the same problem. 
The time preconditioner was not applied in 
Figure~\ref{fig:compm} and Table~\ref{tbl:convt} 
in order to provide a control case against which to compare the effectiveness of the preconditioner.

Figure~\ref{fig:compm} compares the results from different numerical method simulations. Conventional finite difference method (central scheme) is used to compare with FEM. The resolutions $(h=15/5000=0.003)$ are determined near the $r=0$ region. The execution time of the finite difference method (FDM) is slightly slower than the uniform FEM. Results of the FEM with mesh refinement reach the desired residual, and it is faster than the other two methods. These results suggest a benefit in numerical efficiency by using the space-time FEM with mesh refinement. 

\begin{table}[t]
\renewcommand\arraystretch{1.8} 
\captionsetup{justification=RaggedRight}
\begin{center}
\begin{tabular}{|c|c|}
\hline
Tests & Values\\
\hline
$\| \Psi_{FDM} - \Psi_{FEM} \|_2$ & $2.3 \times 10^{-4}$\\
\hline
$(\| \Psi_{h/4} - \Psi_{h/2} \|_2) / (\| \Psi_{h/2} - \Psi_{h} \|_2)$ & 4.712\\
\hline
\end{tabular}
\caption {A comparison of the solution values obtained using the finite
difference method and the space-time FEM with mesh refinement at t=7.99. 
The same conditions as in Figure~\ref{fig:compm} are used in the tests. 
A self convergence test of the space-time FEM results with respect to whole 
the space and time domain is also given, indicating second order convergence.
The $h$ is the same resolution size as in Figure~\ref{fig:solener}.}
\label{tbl:convt}
\end{center}
\end{table}

Table~\ref{tbl:convt} shows the compared solution values of FDM and FEM, and indicates that the solutions of FDM and FEM are extremely close. 
The average value of the self-convergence factor for the FEM is $4.712$,
indicating second order convergence.

\begin{table}[t]
\renewcommand\arraystretch{1.8} 
\captionsetup{justification=RaggedRight}
\begin{center}
\begin{tabular}{|c|c|c|c|}
\hline
$\#$ of subdomains & Iterations & Final Residual & Execution time \\ 
\hline
1 & 4093 &$1.0 \times 10^{-6}$ & $3.9 \times 10^3$sec \\ 
\hline
2 & 3881 &$1.0 \times 10^{-6}$ & $2.8 \times 10^3$sec \\ 
\hline
4 & 2790 &$1.0 \times 10^{-6}$ & $1.0 \times 10^3$sec \\ 
\hline
6 & 2785 &$1.0 \times 10^{-6}$ & $1.0 \times 10^3$sec \\ 
\hline
\end{tabular}
\caption {The performance impact of time-parallel preconditioning. The resolution 
of these tests is $h/10$ ($h =0.003$) with different numbers of time subdomains. The higher resolution $h/10$ is chosen to check effects of time-subdomain numbers. The tests are performed with mesh refinement using the time additive Schwarz preconditioner. The first case presented in the table ($\#$ of subdomains = 1) means that time additive Schwarz preconditioner is not applied to the problem. The maximum iteration number is $5,000$ and $1.0 \times 10^{-6}$ is the desired final residual.}
\label{tbl:compn}
\end{center}
\end{table}

Table~\ref{tbl:compn} compares the results from different time-parallel preconditionings. Results from using different numbers of time-subdomains in
the time decomposition method with refined meshes are 
presented there. Each result reaches the desired residual. Increasing the 
number of time subdomains generally decreases final iteration numbers and execution times. However, once six subdomains are used, the performance improvement
reverses itself. This suggests a limit in amount of concurrency that
the time decomposition approach can ultimately support while opening up a 
new dimension for extracting parallelism in the computation. 

\section{Conclusion}

The semilinear wave equation provides an excellent laboratory for investigating the emergence of singularities from smooth initial data without having to solve the full Einstein equations.  Modeling the formation of singularities and exploring the self-similar solutions that occur near criticality requires highly adaptive methods that are amenable to parallelization. Because of the long time scale of the simulations, time-parallel methods are of particular interest for these studies.  This work has investigated one particular numerical method, a space-time finite element method, that is amenable to time-parallel simulations as well as highly adaptive execution.  Using an additive Schwarz preconditioner in the time direction, this work explored singularity formation in 1+1 and 2+1 dimensions and found self similar solutions in each near criticality.  The impact of the additive Schwarz preconditioner was quantified by comparing time-to-solution with conventional solution approaches, showing an improvement in performance of nearly a factor of four for relatively short time scale runs.  As expected, mesh refinement further enhances this performance improvement.  These performance improvements come at an increased cost in memory usage while providing more opportunities for extraction of parallelism in large time scale simulations.

  There remain many open problems in exploring critical phenomena in 3+1 whose computational costs have hitherto prevented further study.  While creating the four dimensional finite element meshes that would be needed for such calculations is outside the scope of this work, the increased concurrency available from the adaptive time-parallel approach demonstrated here gives a clear performance motivation to explore those open problems using a space-time finite element method.  The ability of space-time finite element methods to enable complicated domain and boundary geometries also suggests some astrophysically relevant application domains, including exploring the neutron star collapse.



\section*{References}
\bibliography{Citation}

\end{document}